\documentclass[aps,twocolumn,superscriptaddress]{revtex4}

\usepackage[dvips]{graphicx}
\usepackage{amssymb}
\usepackage{natbib}
\newcommand{\ped}[1]{\ensuremath{_{\rm #1}}}
\newcommand{\apex}[1]{\ensuremath{^{\rm #1}}}
\begin{document}
\bibliographystyle{apsrev}
\title{Doping and critical-temperature dependence of the energy gaps in Ba(Fe$_{1-x}$Co$_x$)$_2$As$_2$ thin films}
\author{P. Pecchio}
\affiliation{Dipartimento di Scienza Applicata e Tecnologia, Politecnico di Torino, 10129 Italy}
\author{D. Daghero}
\affiliation{Dipartimento di Scienza Applicata e Tecnologia, Politecnico di Torino, 10129 Italy}
\author{G. A. Ummarino}
\affiliation{Dipartimento di Scienza Applicata e Tecnologia, Politecnico di Torino, 10129 Italy}
\author{F. Kurth}
\affiliation{Leibniz-Institut f\"{u}r Festk\"{o}rper-und Werkstoffforschung (IFW) Dresden, P.O.Box 270116, 01171 Dresden, Germany}
\author{B. Holzapfel}
\affiliation{Leibniz-Institut f\"{u}r Festk\"{o}rper-und Werkstoffforschung (IFW) Dresden, P.O.Box 270116, 01171 Dresden, Germany}
\author{K. Iida}
\affiliation{Leibniz-Institut f\"{u}r Festk\"{o}rper-und Werkstoffforschung (IFW) Dresden, P.O.Box 270116, 01171 Dresden, Germany}
\author{R. S. Gonnelli}
\affiliation{Dipartimento di Scienza Applicata e Tecnologia, Politecnico di Torino, 10129 Italy}


\date{\today}

\begin{abstract}
The dependence of the superconducting gaps in epitaxial Ba(Fe$_{1-x}$Co$_{x}$)$_2$As$_2$ thin films on the nominal doping $x$ ($0.04  \leq x \leq 0.15$) was studied by means of point-contact Andreev-reflection spectroscopy. The normalized conductance curves were well fitted by using the 2D Blonder-Tinkham-Klapwijk model with two nodeless, isotropic gaps -- although the possible presence of gap anisotropies cannot be completely excluded. The amplitudes of the two gaps $\Delta\ped{S}$ and $\Delta\ped{L}$ show similar monotonic trends as a function of the local critical temperature $T\ped{c}\apex{A}$ (measured in the same point contacts) from 25 K down to 8 K. The dependence of the gaps on $x$ is well correlated to the trend of the critical temperature, i.e. to the shape of the superconducting region in the phase diagram. When analyzed within a simple three-band Eliashberg model, this trend turns out to be compatible with a mechanism of superconducting coupling mediated by spin fluctuations, whose characteristic energy scales with $T\ped{c}$ according to the empirical law $\Omega_{0}= 4.65 k\ped{B} T\ped{c}$, and with a total electron-boson coupling strength $\lambda\ped{tot}=2.22$ for $x \leq 0.10$ (i.e. up to optimal doping) that slightly decreases to $\lambda\ped{tot}=1.82$ in the overdoped samples ($x = 0.15$).
\end{abstract}
\maketitle

\section{Introduction}
The research on Fe-based superconductors has been recently boosted by the progress in the techniques of film deposition. Films of very high quality are necessary for applications in superconducting electronics, i.e. for the fabrication of Josephson junctions \cite{seidel11}, SQUIDs \cite{katase10b} and so on. However, they can be fruitfully used also to investigate fundamental properties of these compounds. For instance, they are the perfect playground for transport, optical and spectroscopic measurements of various kind; thanks to strain/stress effects that can be induced by the substrate \cite{iida09} thin films offer an additional way to tune the critical temperature; finally, they are necessary to realize some proposed phase-sensitive experiments \cite{golubov13} to determine the order parameter symmetry ($s+\!\!+$ or $s\pm$).

So far, thin films of 122 Fe-based compounds have been used to investigate, for example, the gap amplitude and structure, which are probably the most intriguing open issues of these superconductors. As a matter of fact, the emergence of zeros or nodes in the gap has been predicted theoretically within the $s\pm$ symmetry \cite{kuroki09,graser10,suzuki11,hirschfeld11} as a results of the strong sensitivity of the Fermi surface (FS) to the details of the lattice structure.
In 10\% Co-doped Ba-122 thin films, measurements of the complex dynamical conductivity \cite{fischer10} have shown a small isotropic gap of about 3 meV and a larger, highly anisotropic gap of about 8 meV -- possibly featuring vertical node lines -- located on the electronlike FS sheet. A superconducting gap of 2.8 meV has been measured also by THz conductivity spectroscopy in thin films of the same compound with $T\ped{c}=19$ K, but has been associated to the electronlike FS \cite{nakamura10}. Optical reflectivity and complex transmittivity measurements in Co-doped 122 films (with nominal $x=0.10$) have given instead a isotropic gap of $1.85 \pm 0.15$ meV \cite{gorshunov10}, but have also shown a low-frequency absorption much stronger than expected for a $s$-wave gap. Further measurements of optical conductivity and permittivity in similar films allowed discriminating a small gap $\Delta\ped{S}=1.85$ meV on the electronlike FS and a larger gap $\Delta\ped{L}=3.95$ meV on the holelike FS \cite{maksimov11a}.

Clearly, the results collected up to now do not give a consistent picture, neither about the presence and location of the nodal lines, nor about the amplitude of the gaps. To try to address this point, we have performed point-contact Andreev-reflection spectroscopy (PCARS) measurements in epitaxial Ba(Fe$_{1-x}$Co$_{x}$)$_2$As$_2$ thin films with nominal Co content $x$ ranging from 0.04 to 0.15, i.e. from the underdoped to the overdoped region of the phase diagram. The PCARS spectra do not show any clear hint of the emergence of extended node lines, and can be well fitted by the two-band 2D Blonder-Thinkam-Klapwijk (BTK) model using isotropic gaps -- although the shape of the spectra does not allow excluding some degree of gap anisotropy.
The dependence of the gap amplitudes $\Delta\ped{S}$ and $\Delta\ped{L}$ on the local critical temperature $T\ped{c}\apex{A}$ is discussed. In underdoped and optimally-doped films, the gap ratios are $2\Delta\ped{S}/k\ped{B}T\ped{c}\simeq 3.7$ and $2\Delta\ped{L}/k\ped{B}T\ped{c}\simeq 9$, but decrease to 2.6 and 6.5, respectively, in the overdoped region. When analyzed within a three-band Eliashberg model, these results turn out to be perfectly compatible with $s\pm$ superconductivity mediated by spin fluctuations, whose characteristic energy is $\Omega_0^{sf}=4.65 k\ped{B}T\ped{c}$ (as found experimentally by neutron scattering experiments \cite{inosov10}). An unexpected reduction of the electron-boson coupling strength is observed in the overdoped regime, which could however be reasonably explained by the suppression of spin excitations in this region of the phase diagram.

\section{Experimental details}
%\subsubsection{Epitaxial Grown Thin Films}
The Ba(Fe$_{1-x}$Co$_{x}$)$_2$As$_2$ thin films with a thickness of 50 nm were deposited on $\mathrm{(001) CaF_{2}}$ substrates by pulsed laser deposition (PLD) \cite{kurth13b} using a polycrystalline target with high phase purity \cite{kurth13a,kurth13b}.
The surface smoothness was confirmed by in-situ reflection high energy electron diffraction (RHEED) during the deposition; only streaky pattern were observed for all films indicative of smooth surfaces. The details of the structural characterization and of the microstructure of these high-quality, epitaxial thin films can be found in ref. \cite{kurth13b}. Standard four-probe resistance measurements were performed in a $^4$He cryostat to determine the transport critical temperature and the transition widths, reported in Table \ref{tab:1}. With respect to most phase diagrams of  Ba(Fe$_{1-x}$Co$_{x}$)$_2$As$_2$ single crystals \cite{ni08b,chu09,ning10}, where the optimal doping corresponds to $x\simeq 0.065$, the highest $T\ped{c}^{90}$ of our films is attained for $x=0.10$ and in the $x=0.15$ sample the $T\ped{c}^{90}$ is still about 22 K.
%Whether, and to which extent, this enhancement is due to the mismatch between nominal and effective doping or rather to the effect of the substrate is still under investigation.
This wide doping range of high $T\ped{c}$ is presumably due to a combination of epitaxial strain from the substrate and of reduced Co content in the film with respect to the nominal one. Detailed investigation is underway.
In the following of this paper we will therefore always refer to the doping content of the target. This does not hamper our discussion, since we will refer all the results to the critical temperature of the contact, which is a local property directly correlated to the gap amplitudes (as we have already demonstrated in many different cases \cite{daghero10}) and is thus well defined irrespectively of the actual Co content \footnote{Please note that resistivity measurements unambiguously prove that the film with $x=0.15$ is in the overdoped region, since: i) its $\rho(T)$ curve does not show the low-temperature upturn typical of underdoped samples \cite{chu09}, observed instead in the films with $x=0.04$ and 0.08; ii) its critical temperature is smaller than in the optimally-doped film ($x=0.10$).}.

%\subsubsection{PCARS mesurements}
PCARS measurements have been performed by using the ``soft'' technique, in which a thin Au wire ($\varnothing=18 \,\mu\mathrm{m}$) is kept in contact with the film surface by means of a small drop ($\varnothing \leq 50 \,\mu\mathrm{m}$) of Ag conducting paste. The effective contact size is of course much smaller than the area covered by the Ag paste: parallel nanoscopic contacts are likely to be formed here and there, naturally selecting the more conducting channels within a microscopic region. A possible support to this picture is the fact that, in these films, the standard needle-anvil technique (where a sharp normal tip is gently pressed against the sample surface) does not provide any spectroscopic signal, as experimentally verified by us (note that, however, in films of 1111 compounds the needle-anvil technique has provided good spectroscopic results \cite{naidyuk10}).
The PCARS spectra simply consist of the differential conductance $dI/dV$ of the N-S contact, as a function of the voltage. In principle, a point contact can provide spectroscopic information only if the conduction is ballistic, i.e. electrons do not scatter in the contact region. This is achieved if the contact radius $a$ is smaller than the electronic mean free path, $\ell$ \cite{naidyuklibro}. According to Sharvin's \cite{sharvin65} or Wexler's \cite{wexler66} equations, $a$ is also related to the normal-state contact resistance $R\ped{N}$. The fact that in these films most of the contacts, irrespective of their resistance, do show clear Andreev signals is rather surprising, considering the high residual resistivity (120 $\mu \Omega$ cm for $x=0.10$) of the films, that implies a small mean free path. The precise determination of $\ell$ from the resistivity is not straightforward (at least one should know the plasma frequencies of the different bands, a hard task from the theoretical point of view); however, values of the order of a few nanometers are absolutely reasonable. In these conditions (analogous to those discussed in the case of PCARS on thin films of 1111 compounds \cite{naidyuk10}) the ballistic -- or, at least, the diffusive \cite{naidyuklibro}-- regime can only be achieved when the (microscopic) point contact is the parallel of several nanoscopic contacts that fulfill the ballistic or diffusive conditions, and whose individual resistance is thus much greater that that of the microscopic contact as a whole.

\begin{table}
\begin{tabular}{|c|c|c|c|}
\hline
$x$ & $T\ped{c}\apex{90}$ (K) & $T\ped{c}\apex{10}$ (K) & $\Delta T\ped{c}$ (K) \\
\hline
\hline
0.04 & 9.48 & 7 &  1.24 \\
\hline
0.08 & 23.9 & 25.5 & 0.8 \\
\hline
0.10 & 26.6 & 24.6 & 1.0 \\
\hline
0.15 & 22.0 & 20.6 & 0.7 \\
\hline
\end{tabular}
\caption{Critical temperatures of our films determined from electric transport measurements. $T\ped{c}\apex{90}$ and $T\ped{c}\apex{10}$ are the temperatures at which the resistance (the resistivity) is 90\% and 10\% of the normal-state value immediately before the transition. $\Delta T\ped{c}$ is defined here as $(T\ped{c}\apex{90}-T\ped{c}\apex{10})/2$. The values reported for $x=0.08$ are actually averaged over 3 different films.}\label{tab:1}
\end{table}

Owing to the epitaxial structure of the films and to their surface smoothness, the current that flows through the point contact is mainly parallel to the crystallographic $c$ axis.  By placing the contacts in different regions of the sample surface, we were able to check the homogeneity of the superconducting properties and to obtain some information about their distribution.  To allow a comparison of the experimental $dI/dV$ vs. $V$ curves to the theoretical models, the former must be first normalized, i.e. divided by the normal-state conductance curve $(dI/dV)\ped{N}$ vs. $V$ (in principle, recorder at the same temperature). This curve is inaccessible to experiments because of the very high critical field, and the normal-state conductance measured just above $T\ped{c}$ is unusable because of an anomalous shift of the conductance curves across the superconducting transition, which is typical for very thin films and related to a temperature-dependent spreading resistance contribution (a quantitative explanation of this effect is under study and will be published elsewhere). For these reasons, as shown in detail elsewhere \cite{daghero11}, the normalization can be rather critical in Fe-based compounds; here we chose to divide the low-temperature conductance curves by a polynomial fit of their own high-voltage tails. For the same reason, we will here concentrate on the low-temperature spectra.

The normalized curves were fitted with the BTK model generalized by Kashiwaya and Tanaka \cite{kashiwaya96,kashiwaya00} (later on called ``2D-BTK model'') in order to extract the gap values, as discussed in the following section. This model is based on the assumption of spherical Fermi surfaces (FS) in both the normal metal and the superconductor, which is clearly a strong simplification in the case of Fe-based compounds. A more refined (and complicated) 3D-BTK model could be used to calculate the conductance curves accounting for the real shape of the FS but, as shown elsewhere \cite{gonnelli13,gonnelli13b}, this would not change significantly the resulting gap amplitudes. The local critical temperature (in the following indicated by $T\ped{c}\apex{A}$) can be determined by PCARS by simply looking at the temperature dependence of the raw conductance curves; $T\ped{c}\apex{A}$ is identified with the temperature at which the features related to Andreev reflection disappear and the conductance ceases to be strongly temperature dependent. The $T\ped{c}\apex{A}$ values were generally in a very good agreement with the critical temperature of the whole film determined by resistance measurements and/or susceptibility or magnetization measurements since in most cases they fall between $T\ped{c}^{10}$ and $T\ped{c}^{90}$. A noticeable exception is the underdoped sample ($x=0.04$) in which some spectra show a zero-bias peak that becomes more and more clear on increasing temperature and persists in the normal state. This effect has been observed also by other groups \cite{sheet10,park10,arham13} and might be related to some magnetic scattering rather than to superconductivity. This issue is still under debate \cite{arham13} and will be addressed in a forthcoming paper. Here, however, we will only report PCARS spectra that do not show this anomaly. The absence of $T\ped{c}\apex{A}$ values that fall in the lower $10\%$ of the resistive transition, instead, indicates that there is no significant heating in the contact region (i.e. the temperature of the contact is not significantly higher than that of the bath), as instead would happen if the contacts were in the thermal regime.

\section{Results and discussion}
\begin{figure}[t]
 \includegraphics[width=0.9\columnwidth]{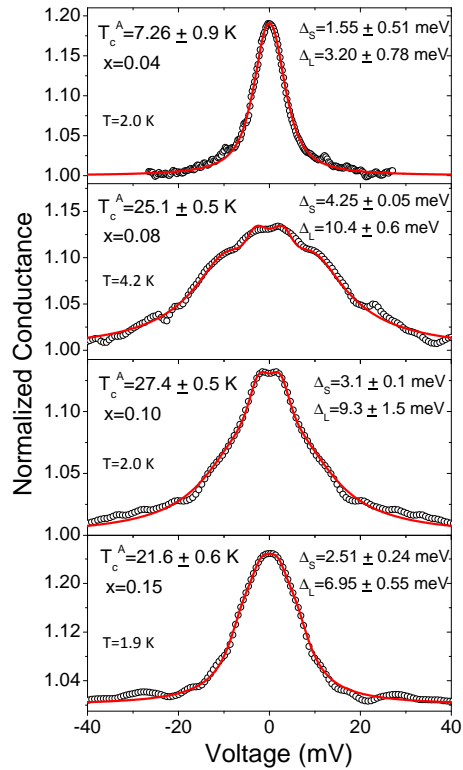}
 \caption{(color online) Low-temperature normalized PCARS spectra in films with different Co content, $x$ (symbols) together with their two-band 2D-BTK fit (lines). The gap values $\Delta\ped{S}$ and $\Delta\ped{L}$ indicated in the legends are actually the average over different possible fits of the same curve, as explained in the text.}\label{fig:spectra}
 \end{figure}

Figure \ref{fig:spectra} shows some representative examples of the many PCARS spectra recorded in films at different doping (symbols), from $x=0.04$ (top panel) to $x=0.15$ (bottom panel). The voltage range on the horizontal axis is the same for all panels so that the variation in the width of the Andreev structures is evident. For $x \geq 0.08$ the shape of all the curves is clearly incompatible with a single gap. These spectra show two symmetric maxima at low energy (or a small flat region around zero bias, as in the bottom panel) which are the hallmark of the small gap $\Delta\ped{S}$, plus additional shoulders or changes in slope at higher energy that are due to the second, larger gap $\Delta\ped{L}$. The case of $x=0.04$, where the double-gap structure is not evident, is in some sense anomalous and will be discussed in more detail later.

Solid lines superimposed to the experimental data represent their best fit within the two-band 2D BTK model. This model assumes that the total conductance is simply the sum of the partial contributions from two sets of equivalent bands, i.e. holelike and electronlike, and each contribution can be calculated by using the 2D BTK model. The model thus contains seven adjustable parameters: the two gap amplitudes $\Delta\ped{S}$ and $\Delta\ped{L}$, the broadening parameters $\Gamma\ped{S}$ and $\Gamma\ped{L}$, the barrier parameters $Z\ped{S}$ and $Z\ped{L}$, and the relative weight of the two bands that contribute to the conductance ($w\ped{S}$ and $w\ped{L}=1-w\ped{S}$). Because of the number of parameters, the set of their best-fitting values for a given spectrum is not univocal, especially when the signal is not very high as in Fe-based compounds. To account for this, we always determined the maximum possible range of $\Delta\ped{S}$ and $\Delta\ped{L}$ values compatible with a given curve, when all the other parameters are changed as well. %By the way, the parameter that mostly affects the ``best fitting'' gap values is the weight $w_1$.
Based on the results obtained in Ba(Fe$_{1-x}$Co$_{x}$)$_2$As$_2$ single crystals at optimal doping \cite{tortello10}, we initially assumed that the two gaps are isotropic. This assumption works well in the \emph{whole} doping range analyzed here, thus indicating that there are no clear signs of  a change in the gap symmetry and structure on increasing the doping content. In this respect it should be noted that the 2D-BTK model is not the most sensitive to the subtle details of the gap structure, so this result does not exclude gap anisotropies either in the plane or in the $c$ direction whose existence has been claimed or predicted in Co-doped Ba-122 \cite{mazin10} and more generally in the 122 systems \cite{graser10,hirschfeld11}. It must be noted, however, that if extended node lines (predicted in particular conditions in 122 compounds \cite{suzuki11}) were present, they would give rise to quasiparticle excitations with very small energy that \emph{can} be detected by PCARS,  as shown in the case of $\mathrm{Ca(Fe,Co)_{2}As_{2}}$ \cite{gonnelli12}.

\begin{figure}[t]
\includegraphics[width=0.9\columnwidth]{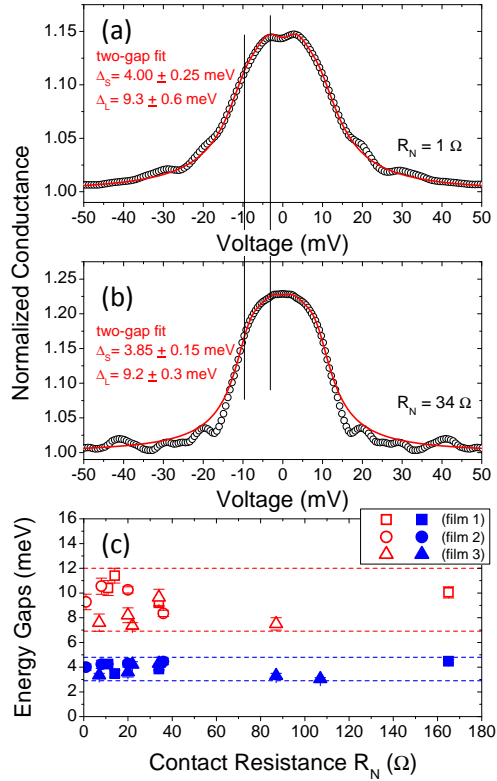}
\caption{(color online) (a, b) Two examples of PCARS spectra taken in different films of $\mathrm{Ba(Fe_{0.92}Co_{0.08})_2As_2}$. Despite the different shape of the spectra, the gap values obtained from the fit are consistent. (c) Gap amplitudes as a function of the resistance of the contacts, which shows the absence of any correlation between these quantities and demonstrates the spectroscopic nature of the contacts. This panel includes data taken in 3 different films. \label{fig:gaps_008}}
\end{figure}

Figure \ref{fig:gaps_008} shows two examples of the many (almost 20) conductance curves measured in films with $x=0.08$. The curves have different shape but the values of the gaps extracted from their fit are compatible with one another.  As shown in panel (c) of the same figure, there is no correlation between the gap values extracted from the fit of different spectra and the resistance of the contact. This fact supports the spectroscopic nature of the contacts \cite{naidyuklibro} and excludes the presence of spreading-resistance effect \cite{chen10b} in our measurements. More generally, the consistency of the gap values obtained in different regions of the \emph{same} film is a good proof of the macroscopic homogeneity of the superconducting properties, while the consistency of the values obtained in \emph{different} films with the same doping is a proof of the reproducibility of the deposition process.

\begin{figure}[t]
\includegraphics[width=0.9\columnwidth]{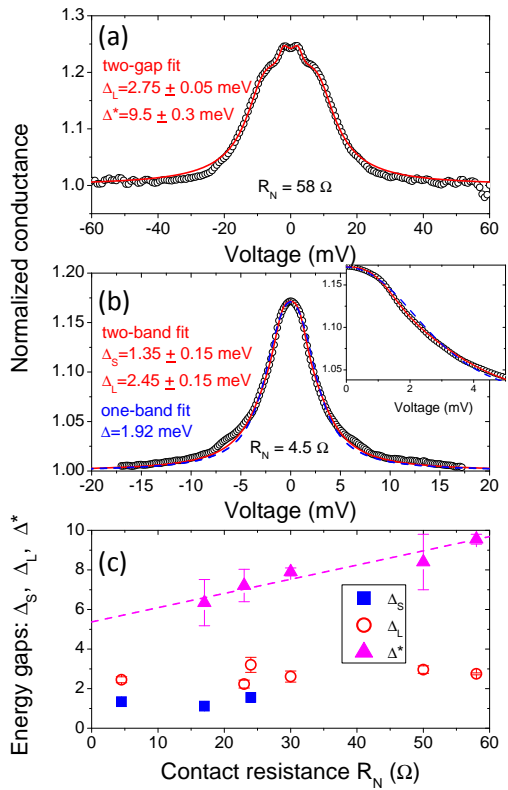}
\caption{(color online) (a, b) Two examples of PCARS spectra taken in different points of the same film ($5 \times 5 \, \mathrm{mm}^2$) of $\mathrm{Ba(Fe_{0.96}Co_{0.04})_2As_2}$. The spectrum in (a) shows clear shoulders around 6 meV and conductance maxima at lower energy. The solid line represents the best-fit of the curve obtained within the 2D-BTK model assuming that the shoulders are due to a superconducting gap $\Delta^*$. The spectrum in (b) instead does not show shoulders but a single maximum at zero bias, and the FWHM of the whole structure is of the order of 3 meV. The solid line is the two-band BTK fit, the dashed line the single-band BTK fit. A magnification of the low-energy region (inset) shows that the two-gap fit is better.  (c) Amplitudes of the ``gap'' $\Delta^*$ and of the gaps $\Delta\ped{S}$ and $\Delta\ped{L}$ as a function of the contact resistance $R\ped{N}$. \label{fig:gaps_004}}
\end{figure}

For $x=0.04$ the spectra often show very clear shoulders at energies of the order of 6 meV in addition to conductance maxima at about 3 meV, as shown in figure \ref{fig:gaps_004}(a). On assumption that the shoulders are due to a superconducting gap, the relevant amplitude (obtained by fitting the curve with the two-band 2D-BTK model) is $\Delta^*=7.9 \pm 1.6 $ meV. Since the measured film showed a $T\ped{c}$ of less than 10 K, this value is clearly unphysical for a superconducting gap. The other gap turns out to be much smaller and ranges between 1.1 meV and 3.2 meV. In a small number of spectra, of which an example is shown in figure \ref{fig:gaps_004}(b),  the structures at about 6 meV are not present and a single, much narrower structure is observed, whose width is of the order of 3 meV. These spectra admit a two-band fit with a small gap $\Delta\ped{S}$ of the order of 1.5 meV and a larger gap $\Delta\ped{L}$ of about 2.5-3.0 meV. Figure \ref{fig:gaps_004}(c) shows a summary of the values of $\Delta\ped{S}$, $\Delta\ped{L}$ and $\Delta^*$ obtained from the two-band fit of spectra of the first and second type, plotted as a function of the resistance of the contacts. Clearly, the larger ``energy scale'' $\Delta^*$ depends on the contact resistance, which indicates that the structures around 6 meV are neither due to a superconducting gap, nor to the strong electron-boson coupling. On the other hand, the smaller gaps do not show a clear dependence on the contact resistance and seem to cluster in two groups indicated by squares and circles for clarity. Although the two energy ranges are very close to each other, they do not overlap (even taking into account the error bars), suggesting that \emph{two} gaps $\Delta\ped{S}$ and $\Delta\ped{L}$ are still present at this doping. Indeed, the few spectra that do not show the high-energy shoulders are better fitted by a two-band model than by a single-band one, as shown in the inset to fig. \ref{fig:gaps_004}(b).

\begin{figure}[t]
 \includegraphics[width=0.9\columnwidth]{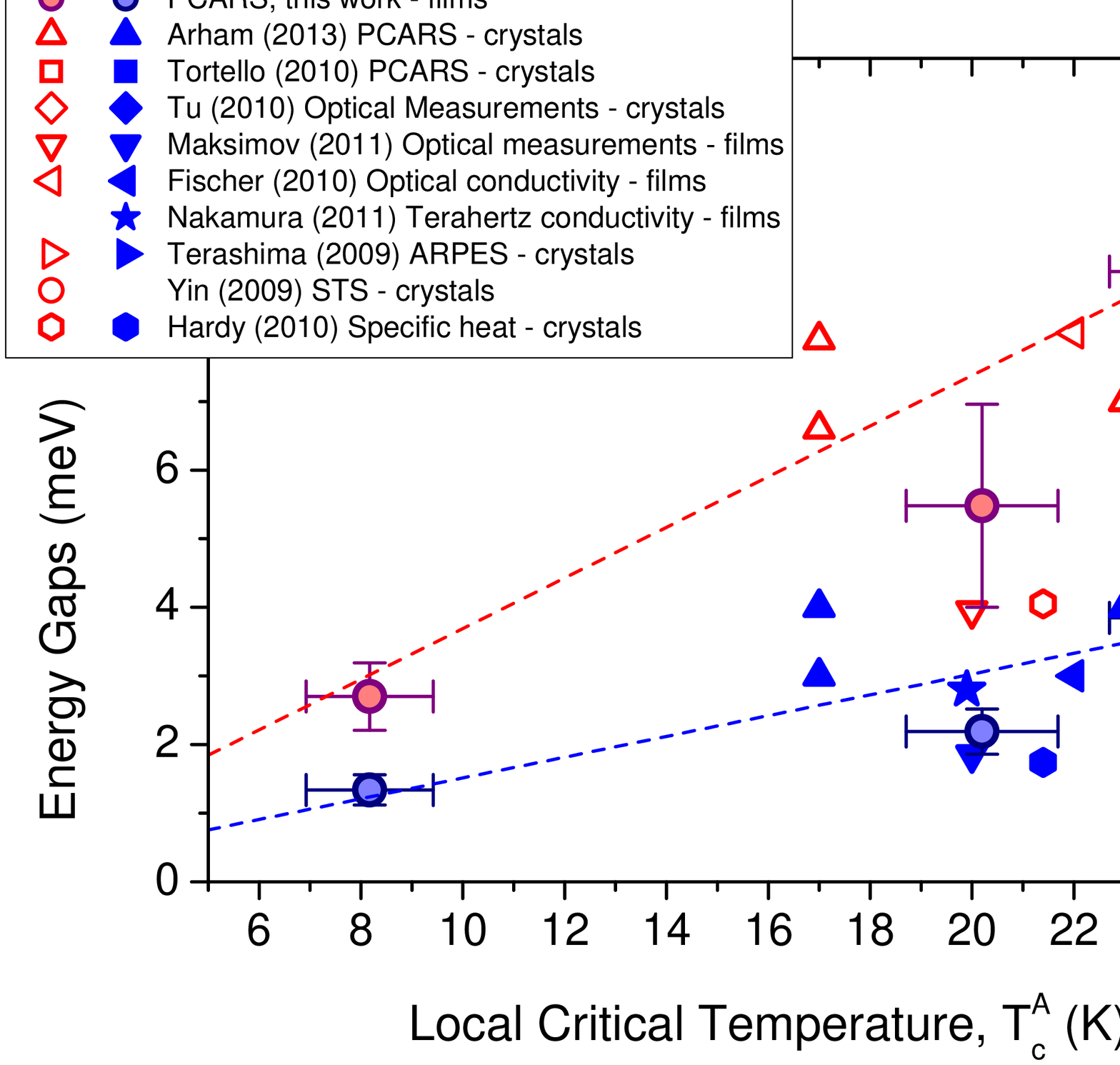}
 \caption{(color online) Average gap amplitudes in Ba(Fe$_{1-x}$Co$_{x}$)$_2$As$_2$ thin films with different Co content obtained by PCARS measurements (filled circles), plotted as a function of $T\ped{c}\apex{A}$. The other data points are taken from literature and specifically: up triangles from \cite{arham13}; squares from \cite{tortello10}; diamonds from \cite{tu10}; down triangles from \cite{maksimov11a}; left triangles from \cite{fischer10}; stars from \cite{nakamura10}; right triangles from \cite{terashima09}; open circles from \cite{yin09}; hexagons from \cite{hardy10}.  The techniques used for these measurements are indicated in the legend. The upper and lower dashed lines correspond to a gap ratio $2\Delta/k\ped{B} T\ped{c}$ equal to 3.52 and 9.0, respectively.}  \label{gaps_vs_Tc}
  \end{figure}

Figure \ref{gaps_vs_Tc} reports the (average) gap amplitudes $\Delta\ped{S}$ and $\Delta\ped{L}$ obtained in the various films as a function of the (average) $T\ped{c}\apex{A}$ of the contacts. In other words, the values of $\Delta\ped{S}$ and $\Delta\ped{L}$ reported here are the midpoints of the corresponding range of gap amplitudes obtained in the fit of different curves. The width of the range is represented by the vertical error bars, while the horizontal error bars indicate the range of $T\ped{c}\apex{A}$ values in all the point contacts made on that film. Figure \ref{gaps_vs_Tc} also shows the results of PCARS in single crystals \cite{tortello10,arham13} as well as the gap amplitudes determined either in films or single crystals by means of other techniques, namely optical measurements \cite{tu10,maksimov11a,fischer10,nakamura10}, specific heat \cite{hardy10}, angle-resolved photoemission spectroscopy (ARPES) \cite{terashima09} and scanning tunneling spectroscopy \cite{yin09}.

\begin{figure}[t]
 \includegraphics[width=0.9\columnwidth]{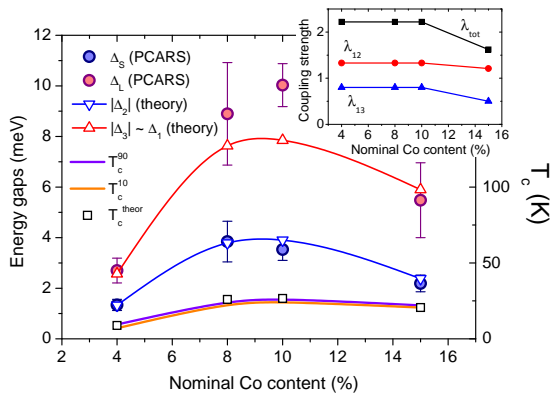}
 \caption{(color online) Doping dependence of the gaps measured by PCARS (circles, left vertical scale) and of the critical temperature from resistivity measurements (lines, right vertical scale). Triangles and squares indicate the values of the gaps and of the critical temperature calculated within the three-band Eliashberg model described in the text. The inset shows the dependence of the coupling strengths $\lambda_{12}$ and $\lambda_{13}$ on the Co content, together with the total electron-boson coupling constant.\label{fig:gaps_vs_x}}
 \end{figure}

At the highest $T\ped{c}$ values, corresponding to $x=0.08$ and $x=0.10$, the gap values agree rather well with those given by PCARS in single crystals \cite{tortello10} and by Arham et al. \cite{arham13}. The large spread of $\Delta\ped{L}$ values given by PCARS has already been noticed in various Fe-based compounds \cite{daghero11} and its origin may be either intrinsic (e.g. anisotropy of $\Delta\ped{L}$) or extrinsic (uncertainty due to the normalization). Finally, the values given by PCARS (especially for $\Delta\ped{L}$) are systematically larger than those given by optical measurements and specific-heat measurements. This may be due to the approximations on which the fit of the curves is based, but may also hide some more fundamental property of Fe-based compounds. The small gap $\Delta\ped{S}$ appears much better defined; the values provided by different techniques are well consistent with one another. Concerning the gap values away from optimal doping, it should be borne in mind that figure \ref{gaps_vs_Tc} reports in the same plot the data in underdoped and overdoped samples; in particular, the points at $T\ped{c} = 20.2$ K refer to the $x=0.15$ film. If these points are temporarily excluded from the analysis, a roughly linear trend of the gaps as a function of $T\ped{c}$ can be observed. The dashed lines in figure \ref{gaps_vs_Tc} have equations $2\Delta/k\ped{B}T\ped{c}=3.52$ and $2\Delta/k\ped{B}T\ped{c}=9.0$; it can be clearly seen that the small gap is approximately BCS for any $x$ between 0.04 and 0.10. Even though $\Delta\ped{L}$ is affected by a much larger uncertainty, it can be said that $2\Delta\ped{L}/k\ped{B}T\ped{c}$ ranges between 7 and 10 in the same doping range. The points at $x=0.15$ are instead outside this trend since the gap values here  correspond to reduced gap ratios.
This point can be clarified by plotting the gap amplitudes as a function of the nominal doping, as in figure \ref{fig:gaps_vs_x}.

As expected, the trend of the gaps mimics the trend of the critical temperature, showing a maximum at $x=0.08-0.10$. However, the trend is not symmetric in the sense that in the overdoped region the gaps decrease ``more'' than the critical temperature, i.e. the gap ratios decrease. The theoretical analysis of these results is presented in the following section.

\section{Interpretation of the results within Eliashberg theory}
%%%%%%%%%%%%%%%%%%%%%%%%%%%%%%%%%%%%%%%%%%%%%%%%%%%%%%%%%%%%%%%%%%%%%%%%%%%%%%%%%%%%%%%%%%%
We have shown elsewhere \cite{ummarino09,ummarino11} that a simple three-band Eliashberg model with a very small number of free parameters can account surprisingly well for the phenomenology of Fe-based superconductors and allows explaining a large variety of their properties.
Here we use the same model to try to rationalize the experimental trend of the gaps as a function of $T\ped{c}$ or of the doping content $x$. The first assumption of the model is that the electronic structure of Ba(Fe$_{1-x}$Co$_{x}$)$_2$As$_2$ can be approximately
described by one hole band (indicated in the
following as band 1) and two electron bands (2 and 3) \cite{tortello10,ummarino11}.
The gap symmetry is assumed to be $s \pm$  \cite{mazin08} so that the sign of $\Delta_{1}$ (here assumed positive) is opposite to that of $\Delta_2$ and $\Delta_3$.
Although PCARS, as well as many other spectroscopic techniques, provides at most two gap amplitudes and does not allow associating them to a particular FS sheet, the use of (at least) three effective bands and thus three gaps is necessary for the Eliashberg model to be able to reproduce the experimental results. However, ARPES results in optimally Co-doped Ba-122 single crystals indicated that the larger gap belongs to the holelike FS sheet \cite{terashima09}. With this in mind, we will assume $\Delta_1 \simeq |\Delta_3|$ and $|\Delta_2|$ to be the large and the small gap measured by PCARS, respectively. This assumption is consistent with the fact that the experimental results do not resolve the two larger gaps.
To obtain the gaps and the critical temperature within the s$\pm$
wave three-band Eliashberg model \cite{eliashberg60} one has to
solve six coupled equations for the gaps $\Delta_{i}(i\omega_{n})$
and the renormalization functions $Z_{i}(i\omega_{n})$, where $i$ is
a band index ($i=1 ..3$). The equations have been reported elsewhere \cite{ummarino09}; their solution requires a large number of input parameters (18 functions and 9 constants); however,
some of these parameters are correlated, some can be
extracted from experiments and some can be fixed by suitable
approximations. For example, the coupling constant matrix $\lambda_{ij}$ can be greatly simplified. In general, one should consider that each matrix element has a contribution from phonons and one from antiferromagnetic (AFM) spin fluctuations (SF), i.e. $\lambda_{ij}=\lambda_{ij}^{ph}+\lambda_{ij}^{sf}$. However, the coupling between the two electron bands is small, and we thus take $\lambda_{23}=\lambda_{32}=0$; the total electron-phonon coupling in pnictides is generally small
\cite{boeri10} and  phonons mainly provide intraband coupling, so that we assume $\lambda_{ij}^{ph}=0$; spin fluctuations mainly provide interband coupling between the two quasi-nested FS sheets \cite{mazin08}, and thus we assume $\lambda_{ii}^{sf}=0$. Finally, the electron-boson
coupling-constant matrix $\lambda_{ij}$ takes the following form:
\cite{mazin09,ummarino09,tortello10}:
\begin{equation}
\vspace{2mm} %
\lambda_{ij}= \left (
\begin{array}{ccc}
  \lambda^{ph}_{11}             &             \lambda^{sf}_{12}                  &               \lambda^{sf}_{13}            \\
  \lambda^{sf}_{21}                &               \lambda^{ph}_{22}               &               0            \\
  \lambda^{sf}_{31}             &  0   & \lambda^{ph}_{33} \\
\end{array}
\right ) \label{eq:matrix}
\end{equation}
where $\lambda^{sf}_{21}=\lambda^{sf}_{12}\nu_{12}$ and $\lambda^{sf}_{31}=\lambda^{sf}_{13}\nu_{13}$,
with $\nu_{ij}=N_{i}(0)/N_{j}(0)$ and $N_{i}(0)$ is the normal
density of states at the Fermi level for the $i$-th band.
Another fundamental ingredient is the electron-boson spectral function $\alpha^2 F(\Omega)$ of the boson responsible for the pairing. The shape of the electron-phonon spectral function is taken from literature \cite{mittal08} and we assume $\alpha_{11}^2F^{ph}_{11}(\Omega)=\alpha_{22}^2F^{ph}_{22}(\Omega)=\alpha_{33}^2F^{ph}_{33}(\Omega)$
with $\lambda_{ii}^{ph}=0.2$ \cite{popovich10}.
As for spin fluctuations, we assume their spectrum to have a Lorentzian shape \cite{ummarino09,ummarino11b,ummarino12,daghero12}:
\begin{equation}
\alpha_{ij}^2F^{sp}_{ij}(\Omega)=
C_{ij}\big\{L(\Omega+\Omega_{ij},Y_{ij})-
L(\Omega-\Omega_{ij},Y_{ij})\big\}
\end{equation}
where $L(\Omega\pm\Omega_{ij},Y_{ij})=\frac{1}{(\Omega
\pm\Omega_{ij})^2+Y_{ij}^2}$ and $C_{ij}$ are normalization
constants, necessary to obtain the proper values of $\lambda_{ij}$
while $\Omega_{ij}$ and $Y_{ij}$ are the peak energies and
half-widths of the Lorentzian functions, respectively \cite{ummarino09}.
In all the calculations we set
$\Omega_{ij}=\Omega_0^{sf}$
and $Y_{ij}=Y_{ij}^{sf} = \Omega_{0}^{sf}/2$ \cite{inosov10}.
Here, $\Omega_{0}^{sf}$ is the characteristic energy  of the AFM SF, assumed to be equal to the spin-resonance energy, as verified experimentally by us in optimally Co-doped Ba-122 single crystals \cite{tortello10,daghero11}. Its value is determined according to the empirical relation $\Omega_{0}^{sf}=4.65 k\ped{B}T\ped{c}$ (proposed in ref. \cite{paglione10}) holds. Bandstructure calculations provide information about the factors $\nu_{ij}$ that enter the definition of
$\lambda_{ij}$. In the case of optimally-doped Ba(Fe$_{1-x}$Co$_{x}$)$_2$As$_2$, $\nu_{12}=1.12$ and
$\nu_{13}=4.50$ \cite{mazin_private}. As a first approximation, these values have been used here for all Co contents.
Moreover, we assume for simplicity that all the elements of the Coulomb pseudopotential matrix are identically zero ($\mu^{*}_{ii}=\mu^{*}_{ij}=0$), and we neglect the effect of disorder, owing to the high quality of the films.

Finally, only two free parameters remain, i.e. the coupling constants $\lambda_{12}^{sf}$ and $\lambda_{13}^{sf}$. These parameters can be tuned in such a way to reproduce the experimental values of the small gap $\Delta\ped{S}$ and of the critical temperature, which are the best-defined experimental data; the values of the large gap $\Delta\ped{L}$ are indeed affected by a larger relative uncertainty, and moreover they might actually be a sort of weighted ``average'' of the two gaps $\Delta_1$ and $|\Delta_3|$. The larger gaps are therefore calculated with the values of $\lambda^{sf}_{12}$ and $\lambda^{sf}_{13}$ that allow reproducing $\Delta\ped{S}$ and $T\ped{c}$.

The result of these calculations is that: i) the trend of the experimental gaps $\Delta\ped{S}$ and $\Delta\ped{L}$ as a function of $T\ped{c}$ and of $x$ in the samples with nominal Co content $x=0.04, \, 0.08$ and $0.10$ can be reproduced by using $\lambda_{12}^{sf} =0.8$ and $\lambda_{13}^{sf}=1.33$, and only changing the value of the characteristic SF energy $\Omega_0$ according to the change in $T\ped{c}$; ii) to reproduce the values of the gaps and of $T\ped{c}$ in the overdoped sample ($x=0.15$) it is instead also necessary to reduce the values of the two coupling constant: $\lambda_{12}^{sf}=0.5$ and $\lambda_{13}^{sf}= 1.21$. The values of these two parameters are shown as a function of $x$ in the inset of figure \ref{fig:gaps_vs_x}. Note that the total coupling is $\lambda\ped{tot}=2.22$ for $x=0.04, \, 0.08$ and $0.10$ and decreases to $\lambda\ped{tot}= 1.82$ at $x=0.15$. These values are in agreement with those found in previous works \cite{ummarino11,popovich10}, and indicate that Co-doped Ba-122 is a strong-coupling superconductor at all the doping contents analyzed here.
The main panel of figure \ref{fig:gaps_vs_x} also reports the calculated values of the gaps as a function of $x$. The agreement between the theoretical and experimental values of $T\ped{c}$ and of the small gap is very good; the large gap is underestimated around optimal doping, but the trend is qualitatively correct. The agreement might be improved if the feedback effect of the condensate on the bosonic excitations \cite{ummarino12,daghero12} was taken into account, which was not done in this paper for simplicity.

\section{Conclusions}
In conclusion, we have determined the energy gaps of Ba(Fe$_{1-x}$Co$_x$)$_2$As$_2$ in a wide range of nominal doping ($0.04 \leq x \leq 0.15$) by means of ``soft"  PCARS measurements in epitaxial thin films.  Several PCARS spectra were acquired on each sample, with the probe current injected perpendicular to the film surface and thus mainly along the $c$ axis. In all films, the PCARS spectra admit a fit with the two-band 2D-BTK model using two isotropic gaps, and their shape does not suggest the presence of node lines on the FS. This means that superconductivity in this system keeps its multiband character even when the critical temperature is of the order of 10 K, and that there are no clear hints of changes in the gap symmetry or structure in the doping range of our films -- although the shape of the spectra does not allow excluding some degree of gap anisotropy.

The small gap turns out to be approximately BCS, with a ratio $2\Delta\ped{S}/k\ped{B}T\ped{c}=3.7 \pm 0.8$ (the uncertainty arises from the statistical spread of gap values) for $x\leq 0.10$, and smaller ($2.6 \pm 0.3$) at $x=0.15$. The second gap is much larger, with a ratio $2\Delta\ped{L}/k\ped{B}T\ped{c}$ of the order of 9 for $x\leq 0.10$ and 6.5 for $x=0.15$.

The trend of the gaps and of $T\ped{c}$ as a function of the Co content can be reproduced by a simple $s\pm$ Eliashberg model in which the spectrum of the mediating boson is that of spin fluctuations, and its characteristic energy coincides with the energy of the spin resonance. The decrease of the gap ratios in the overdoped samples is reflected in the values of the coupling strengths that are constant for $x \leq 0.10$ and slightly decrease at $x=0.15$. This result finds a natural explanation within the picture of $s\pm$ superconductivity mediated by spin fluctuations: in the overdoped regime, far from the AFM region of the phase diagram, superconductivity may suffer from a suppression of the spin fluctuations and the loss of nesting \cite{fang09}, which could lead to a decrease in the superconducting interband coupling that, in turns, produces a larger decrease of the gaps in comparison with the reduction of the critical temperature.

D.D. and P.P. wish to thank the Leibniz Institute for Solid State and Materials Research (IFW) in Dresden, Germany, and in particular the Department of Superconducting Materials, where many of the PCARS measurements were performed. Particular thanks to V. Grinenko, J. H\"{a}nisch and K. Nenkov for valuable discussions and technical support.

This work was done under the Collaborative EU-Japan Project ``IRON SEA'' (NMP3-SL-2011-283141).

\bibliography{bibliografia_defBa122copia2}
\newpage

\end{document}